# Long Range Charge Ordering in Magnetite Below the Verwey Transition.

## J. P. Wright,[1][†] J. P. Attfield,[1][*] and P. G. Radaelli[2]

[1]Department of Chemistry, University of Cambridge, Lensfield Road, Cambridge CB2 1EW, U.K.

[2]ISIS Facility, Rutherford Appleton Laboratory, Chilton, Didcot, OX11 0QX, U.K.

*To whom correspondence should be addressed. E-mail: jpa14@cam.ac.uk

†New address: ESRF, BP 220, Grenoble, France.

**The crystal structure of $Fe_3O_4$ below the 122 K Verwey transition has been refined using high-resolution X-ray and neutron powder diffraction data. The refinements give direct evidence for charge ordering (CO) over four independent octahedral Fe sites, two with a charge of +2.4 and the other two of +2.6. CO schemes consistent with our model do not meet the widely-accepted Anderson condition of minimum electrostatic repulsion. Instead we propose that CO is driven primarily by a [001] electronic instability, which opens a gap at the transition through a charge density wave mechanism.**

Mechanisms for the spatial ordering of different valence states in transition metal oxides are of current interest, as charge-ordered 'stripes' have been evidenced in many manganite perovskites [1,2] and superconducting cuprates [3], and their dynamic fluctuations have been proposed as a mechanism of high temperature superconductivity [4]. The classic charge ordering problem in transition metal oxides is that of magnetite, which has been unresolved for over 60 years, and which we have reinvestigated through this structural study.

Magnetite is the parent compound for magnetic materials such as $\gamma$-$Fe_2O_3$ and the spinel ferrites, and its magnetoresistive properties have been of recent interest for spin electronics [5]. The inverse spinel crystal structure is written as $Fe^{3+}[Fe^{2+}Fe^{3+}]O_4$ to distinguish the first (A type) tetrahedrally-coordinated $Fe^{3+}$ from the bracketed $Fe^{2+}$ and $Fe^{3+}$ ions in (B type) octahedral sites. The two B cation magnetic moments are antiparallel to the single A cation moment resulting in ferrimagnetism, and rapid electron hopping between the B site $Fe^{2+}$ and $Fe^{3+}$ ions renders these sites structurally and spectroscopically equivalent, and gives rise to the moderate electronic conductivity (4 m$\Omega$.cm at 300 K).

Verwey [6] discovered that magnetite undergoes a sharp, first order transition on cooling below ~120 K at which the resistivity of magnetite increases sharply by two orders of magnitude, and the structure distorts from cubic symmetry [7]. This was explained by a charge ordering (CO) of the $Fe^{2+}$ and $Fe^{3+}$ states on the B sites in alternating layers, although this was not confirmed despite much initial research [8]. Mossbauer spectra demonstrate that localised $Fe^{2+}$ and $Fe^{3+}$ are present at the octahedral sites, and a recent study resolved two environments for each cation in addition to the tetrahedral $Fe^{3+}$ signal [9]. The low temperature structure is a rhombohedral distortion of the cubic spinel arrangement to the first approximation [10] but this splits the B sites in a 3:1 ratio and so is not consistent with the expected CO. Neutron [11] and X-ray [12] diffraction studies of single crystals and electron microscopy [13] have shown that the low temperature structure has a monoclinic $\sqrt{2}a$ x $\sqrt{2}a$ x 2a supercell (*a* is the cell parameter of the undistorted cubic phase) with Cc space group symmetry, containing 8 unique A type and 16 unique B type cation sites. A recent $^{57}$Fe NMR study [14] has resolved the 8 tetrahedral and 15 of the 16 octahedral iron environments. The only published refinement of the low temperature superstructure of magnetite was based on a single crystal neutron diffraction experiment at 10 K by Iizumi *et al* [15]. An $a/\sqrt{2}$ x $a\sqrt{2}$ x 2*a* subcell was used with additional Pmca or Pmc2$_1$ orthorhombic symmetry constraints to reduce the complexity of the model. Although some differences between the inequivalent B sites were found, these were not concluded to be significant evidence for CO. A recent resonant X-ray diffraction study also reported no evidence for charge disproportionation above a sensitivity limit of 25% [16].

As well as being an inherently complex problem, the low temperature structure of magnetite is difficult to study by single crystal diffraction as severe twinning occurs below the Verwey transition. Mechanical detwinning and magnetic alignment were used by Iizumi *et al* [15], but multiple scattering and extinction problems were still encountered. To overcome these difficulties, we have performed a combined powder X-ray and neutron diffraction study, using very high resolution instruments at intense synchrotron X-ray and pulsed neutron sources.

A highly stoichiometric sample of powdered magnetite ($Fe_{3-\delta}O_4$ with $\delta < 0.0001$) was obtained by grinding single crystals grown by the skull melter technique [17]. High resolution powder diffraction data were collected at 90 and 130 K. Neutron time-of-flight data were obtained from instrument HRPD at the ISIS spallation neutron facility, UK, for approximately 25 hours at each temperature in the range d = 0.31 – 4.45 Å. The data were focussed and normalised using a standard vanadium spectrum. Synchrotron x-ray data were collected on the BM16 instrument at ESRF, Grenoble, France for approximately 18 hours at each temperature. The wavelength was 0.49395 Å and data were collected between $2\theta = 0$ and 70º (minimum d = 0.43 Å) from a 0.7 mm capillary sample. Rapid, variable temperature x-ray scans located the Verwey transition for this sample at 122 K, confirming that it is oxygen stoichiometric.

The 130 K diffraction patterns contain the peaks expected from cubic magnetite. Weak contributions from 0.8 wt. % $\alpha$-$Fe_2O_3$, which may have resulted from surface oxidation of the original crystals, and some scattering from aluminium in the sample environment in the neutron patterns were included in the Rietveld fits to the data. The magnetic diffraction intensities were also fitted in the neutron profile. The 90 K patterns show many splittings of the principal magnetite peaks. Weak superstructure peaks with <1% of the intensity of the principal peaks are also observed (Fig. 1). These can be indexed on the monoclinic $a/\sqrt{2}$ x $a\sqrt{2}$ x $2a$ subcell with P2/c or Pc symmetry, except for three very weak peaks which require the full $\sqrt{2}a$ x $\sqrt{2}a$ x $2a$ Cc supercell. Hence, structure refinement was only attempted on the former subcell [18]. Refinement in the centric monoclinic space group P2/c was unstable, and so symmetry constraints from the orthorhombic Pmca group were applied [15]. This is equivalent to averaging the true superstructure over the additional symmetry operators since the reflections not indexed by this model are extremely weak. A structural model was generated by randomly displacing the atoms from their idealised cubic magnetite positions. Separate refinements of this model with the neutron and synchrotron profiles converged successfully giving similar values for the atomic coordinates. Refinements using alternative orthorhombic symmetry constraints gave poorer or unstable fits. The final model was fitted to the X-ray and neutron data simultaneously in the range 0.5 < d < 4.0 Å. The contrast between the Fe:O scattering factor ratio of ~3 for X-rays and 1.6 for neutrons reduces correlations between the refined Fe and O coordinates. The residuals were $R_{WP}$ = 6.81%, $R_P$ = 4.78 % and reduced-$\chi^2$ = 5.92 and the refined cell parameters were a = 5.94437(1) Å, b = 5.92471(2) Å, c = 16.77512(4) Å and β = 90.236(1)°. Refined coordinates and other results will be published elsewhere.

Our refinement confirms the correctness of the symmetry approximations used by Iizumi *et al* [15], however, the refined parameters differ from those in the latter study, leading to different values for the Fe-O bond distances. These are sensitive experimental indicators of the iron charge state. The expected average Fe-O distances for high spin octahedral $Fe^{3+}$ and $Fe^{2+}$ are 2.025 and 2.160 Å, respectively [19], so within the experimental errors of <0.005 Å (Table I) the refined model is sensitive to >4% charge differences between the Fe sites. The average Fe-O distances in Table I are as expected for the tetrahedral sites; those for the octahedral cations fall into two groups, with B2 and B3 being significantly smaller than B1 and B4. We stress that the coincidences of these pairs of average Fe-O distances are not an artifact of the symmetry constraints in the refinement.

Bond valence sums (BVS's) are a more quantitative indicator of overall charge state and were calculated using a standard method and parameters [20]. The BVS's for the six oxygen sites were between 1.94 and 2.06 and those for the Fe cations are shown in Table I. Estimated site valences V were also obtained by constraining the A site charges to be +3 and renormalising the BVS values to average the B site charges to +2.5. The BVS and V values for the B sites fall into two clear groups with a charge difference of 0.2e between the large (B1 and B4) and small (B2 and B3) sites. Thus, even with symmetry-averaging orthorhombic constraints, this refinement clearly demonstrates that the octahedral Fe sites in magnetite are split into two groups with an estimated charge difference of 20% of that expected for ideal $Fe^{2+}$ and $Fe^{3+}$ states. This provides direct crystallographic evidence of at least partial long range charge ordering on the B sites below the Verwey transition.

Each B site in our model is averaged over four inequivalent subsites in the larger $\sqrt{2}a$ x $\sqrt{2}a$ x 2a Cc supercell, so that our large (B1 and B4) and small (B2 and B3) subsites could respectively be averaged over $4Fe^{2+}$ and $4Fe^{3+}$ subsites (which we refer to as class I CO), or over $(3Fe^{2+} + Fe^{3+})$ and $(Fe^{2+} + 3Fe^{3+})$ subsites (class II CO). A recent comparison [21] of four unfrustrated, CO structures in transition metal oxide perovskites such as $Ca_2Fe^{3+}Fe^{5+}O_6$ and $YBaCo^{2+}Co^{3+}O_5$ has shown that the observed charge difference based on BVS's is only 20-60% of that expected for the ideal valence states. The observed charge difference in magnetite is 20% (for class I CO) or 40% (for class II) of the expected value and therefore is not anomalous in comparison to other oxides. It is hard to ascertain whether this general reduction in the degree of charge ordering results from non-localised electron density (fractional charge ordering), strong mixing of ground and excited states, strained charge states, inter-site disorder, or a combination of these effects.

The simplest description of the CO in magnetite is as the ordering of $N/2$ electrons on a lattice of $N$ B-type $Fe^{3+}$ cations. This CO is highly frustrated as the B sites are arranged in a network of corner-sharing tetrahedra so that it is impossible for electrons to avoid occupying adjacent sites. The nearest neighbour electrostatic energy per electron $E_{nn}$ is minimised at $U = e^2/d_{Fe-Fe}$ ($d_{Fe-Fe}$ is the shortest Fe-Fe distance) when each tetrahedron contains two electrons, i.e. two $Fe^{2+}$ and two $Fe^{3+}$ ions. This minimum energy condition for any long or short range CO was proposed by Anderson [22] and has been treated as a prerequisite in subsequent investigations [23].

In our Pmca-symmetry averaged model, the Anderson condition requires the B1 and B2 sites to have the same average charge 2.5+q, while B3 and B4 should have the complementary average charge 2.5-q. Our observation that sites B1 and B4 have the same average charge of +2.4 while B2 and B3 have the complementary value shows that the Anderson condition is not satisfied, so CO models that are consistent with our refinement have higher $E_{nn}$ energies. Class I CO represents a single charge ordered model in the Cc supercell, in which each tetrahedron of B sites contains one or three electrons giving $E_{nn} = 1.5U$. There are 32 fully charge ordered models in class II, 9 of which have a minimum $E_{nn} = 1.25U$. One of these arrangements (Figure 2) minimises repulsions between higher neighbours and we propose this as a plausible model for long range charge ordering in magnetite that is consistent with the diffraction results. However, our results do not rule out the other class I and II models, nor do they show whether the charges are fully long range ordered on any or all of the sites B1-B4. Distinguishing between these possibilities would require reliable structure factors for the Cc superstructure reflections.

The observation that the Anderson condition of minimum electostatic repulsion is not met in the low temperature structure of $Fe_3O_4$ is not surprising when compared to other CO structures.

Although some arrangements are consistent with a minimum repulsion between charges, e.g. in $LaBaMn^{2+}Mn^{3+}O_5$ [24] and the warwickite $Fe^{2+}Fe^{3+}OBO_3$ [25], isostructural materials show alternative orderings, e.g. in $YBaCo^{2+}Co^{3+}O_5$ [26] and $Mn^{2+}Mn^{3+}OBO_3$ [27], in which the CO is driven by local structural distortions associated with the charge states such as orbital ordering, and an Anderson-type criterion is not fulfilled. The role of local distortions in the low temperature magnetite structure is hard to assess as the full superstructure remains unknown, but these could be significant as the $Fe^{2+}$ is orbitally degenerate. A magnetostriction caused by the localisation of $Fe^{2+}$ may lead to the rhombohedral part of the distortion at the Verwey transition.

An important feature in our refinement is that both the displacements of the atoms from their ideal cubic spinel positions and the B site charges follow a pronounced [001] wave (indexed on the cubic cell). The [001] atomic displacement wave of amplitude ~0.05 Å was noted previously [15] and has been confirmed by recent diffuse scattering experiments [28], but a [001] charge density wave (CDW) has not been reported before and is incompatible with the Anderson criterion. Recent band structure calculations for cubic magnetite have revealed an [001] nesting vector at the Fermi surface in the B-site minority spin electron band [29]. We propose that this instability is relieved via a CDW mechanism [30], in which the small distortions associated with the charge ordering create a gap in the energy band structure leading to the observed loss of conductivity at the Verwey transition. Thus charge ordering in magnetite is driven primarily by the distortions that create a gap in the single-electron energy spectrum, and not by electron-electron repulsions or orbital ordering, although these may play a secondary role in giving rise to the additional $[00^1/_2]$ structural modulation.

We conclude that significant (20%) charge ordering is present in magnetite below the Verwey transition and that this does not satisfy the Anderson criterion of minimum electrostatic energy. Both of these observations are consistent with other, recently determined, charge ordered transition metal oxide structures. The charge ordering has a pronounced [001] modulation which is consistent with the gap that opens at the Verwey transition, although other structural modulations such as $[00^1/_2]$ also result. A fully charge ordered structure with Cc or lower symmetry remains to be determined.

We thank Prof. J.M. Honig (Purdue University) for supplying the magnetite sample and for his advice and encouragement, Dr. A. Fitch for assistance with data collection at ESRF, and EPSRC for providing beamtime and a studentship for JPW.

**Figure Captions**

Figure 1. Part of the high resolution powder X-ray (upper panels) and neutron (lower panels) diffraction patterns of $Fe_3O_4$. The intensity scales are logarithmic in order to emphasise the weak superstructure peaks. Observed (crosses), calculated (full lines) and difference (as difference/estimated standard deviation) plots are shown for the fit to the 90 K data. The observed patterns at 130 K (above the Verwey transition) are also plotted one decade above the 90 K data. Markers show the positions of the Bragg reflections in the low temperature $Fe_3O_4$ structure, markers for aluminium in the sample environment are also shown for the neutron data.

Figure 2. A model for charge ordering on the B sites in the $\sqrt{2}a$ x $\sqrt{2}a$ x $2a$ Cc supercell of magnetite that is consistent with the class II solutions from our refinement. The B sites are numbered as in Table I. Dark/light circles correspond to $Fe^{2+}/Fe^{3+}$. The charge per 4 B sites relative to the average value of +10 is shown for each layer; these define a [001] charge density wave.

Table I Results for the distinct tetrahedral (A1,A2) and octahedral (B1-B4) Fe sites in the refined structure of $Fe_3O_4$ at 90 K; individual and mean distances to the coordinating oxygens (O1-O6), Bond Valence Sums (BVS), and the renormalised valences V.

| Fe site | -O bonds | d(Fe-O)/ Å | <d(Fe-O)>/ Å | BVS | V |
|---------|----------|-----------|-------------|-----|---|
| A1 | -O1 | 1.898(4) | 1.886(3) | 2.80 | 3.00 |
|    | -O5 | 1.882(2) |          |      |      |
|    | -O5 | 1.875(2) |          |      |      |
|    | -O2 | 1.890(4) |          |      |      |
| A2 | -O4 | 1.913(5) | 1.890(4) | 2.77 | 3.00 |
|    | -O6 | 1.877(2) |          |      |      |
|    | -O6 | 1.870(2) |          |      |      |
|    | -O3 | 1.899(5) |          |      |      |
| B1 | -O1 (x2) | 2.042(3) | 2.072(3) | 2.50 | 2.39 |
|    | -O2 (x2) | 2.082(3) |          |      |      |
|    | -O6 (x2) | 2.091(3) |          |      |      |
| B2 | -O3 (x2) | 2.038(4) | 2.043(3) | 2.73 | 2.61 |
|    | -O4 (x2) | 2.040(3) |          |      |      |
|    | -O5 (x2) | 2.052(3) |          |      |      |
| B3 | -O2 | 1.964(5) | 2.050(4) | 2.71 | 2.59 |
|    | -O3 | 2.116(5) |          |      |      |
|    | -O5 | 2.092(3) |          |      |      |
|    | -O5 | 2.091(3) |          |      |      |
|    | -O6 | 2.019(3) |          |      |      |
|    | -O6 | 2.018(3) |          |      |      |
| B4 | -O1 | 2.033(5) | 2.069(4) | 2.52 | 2.41 |
|    | -O4 | 2.086(5) |          |      |      |
|    | -O5 | 2.094(3) |          |      |      |
|    | -O5 | 2.093(3) |          |      |      |
|    | -O6 | 2.054(3) |          |      |      |
|    | -O6 | 2.053(3) |          |      |      |

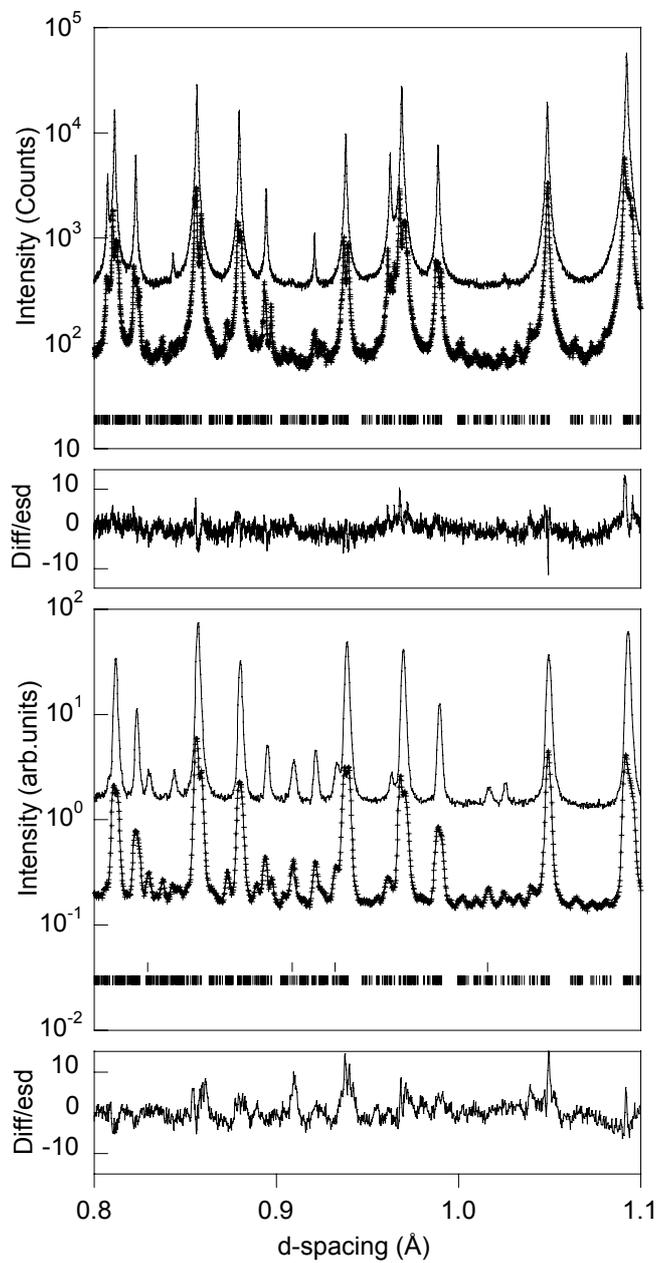

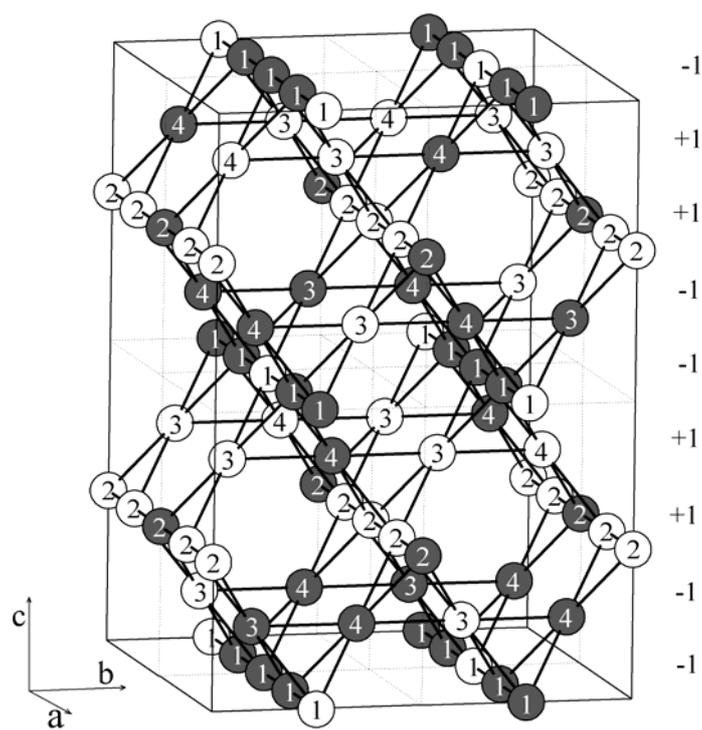

-1
+1
+1
-1
-1
+1
+1
-1
-1

**J. P. Wright** *et al.*
**Figure 2**